\documentstyle[12pt]{article}
\def\pbar{\overline{p}}
\begin{document}
\baselineskip=14pt
\title{\normalsize \bf  CUMULANT TO FACTORIAL MOMENT RATIO \\
	AND MULTIPLICITY DISTRIBUTIONS }
\author{{\small \sc GABRIELE~GIANINI}\thanks{
co-authors:
I.~M.~Dremin, V.A. Nechitailo - P.N.~Lebedev Institute of Physics, Moscow;
B.~B.~Levtchenko - Moscow State University;
V.~Arena, G.~Boca, S.~Malvezzi, M.~Merlo, S.P.~Ratti,
C.~Riccardi, G.~Salvadori, L.~Viola, P.~Vitulo - University of Pavia and INFN.
}\\
{ \small \it Department of Physics, University of Pavia}\\
{ \small \it and INFN, via Bassi 6, Pavia I-27100, Italy }}
\date{}
\maketitle
\begin{abstract}
    The ratio of cumulant to factorial moments of multiplicity
distributions has been calculated for $e^{+}e^{-}$ and $hh$ data
in a wide range of energies. As a function of the rank the ratio exhibits
a regular behaviour with a steep descent and two negative minima.
\end{abstract}

\vspace{1pt}
\begin{flushleft}
{\bf \large 1. Introduction}
\end{flushleft}
\vspace{1pt}

Probing QCD predictions and phenomenological
hypotheses upon the shape of multiplicity  distributions
is a major issue in Multiparticle Dynamics, and requires
a careful comparison between experimental data and theory.
Usually this task is accomplished through the
study of the shape parameters of the distribution
or through the examination of the behaviour
of the factorial moments as a function of their rank.
However, in a recent paper$^{1a}$
the possibility has been suggested of investigating
multiplicity distributions through the ratio:
\(
H_{q} = K_{q}/F_{q}
\)
of cumulant and factorial moments.

The $q$-th order factorial moments $F_{q}$ and cumulant moments $K_{q}$
can be easily obtained from any probability distribution $P_{n}$
by means of the following relations
\[
F_{q}= {\Sigma_{n=1}^{\infty} n(n-1)...(n-q+1)P_{n}/ <n>^{q} },
\;\;\;\;\;\;\;\;\;\;\;
K_{q}= { F_{q} - \Sigma_{m=1}^{q-1} C_{q-1}^{m} \; K_{q-m} \; F_{m} }\;,
\]
with $F_0=F_1 =K_1 =1, K_0 =0$ and where $C_{q-1}^{m}=q!/m!(q-m)!$.
For the most common theoretical multiplicity distributions,
the $F_{q}$'s are  characterized by absolute values rapidly increasing with
the rank$^{1b}$ and display patterns that are not easy to distinguish
from one another;
the ratios \(H_{q}\), on the contrary, under the same
conditions lead to easily distinguishable patterns and
remain bounded$^{1b,1c}$.
For instance:
Negative Binomial Distribution (NBD)
leads to positive defined $H_{q}$'s,
monotonically decreasing with the rank$^{1c}$;
similar features are expected by the QCD parton
\begin{table}[t]
\caption{Investigated data.}
\begin{center}
\begin{tabular}{||l|r||l|r||}
\hline
% & & & \\
$e^{+}e^{-}$  & $\sqrt{s}$ & $hh$  & $\sqrt{s}$ \\
% & & & \\
\hline
% & & & \\
TASSO$^{2a}$ 	&   22 GeV 	& $pp$ 30" B.C. at FNAL$^{3a}$ & 23.8 GeV \\
HRS$^{2b}$   	&   29 GeV 	& $pp$ SMF at CERN$^{3b}$ & 30.4 GeV \\
TASSO$^{2a}$     & 34.8 GeV 	& $pp$ 30" B.C. at FNAL$^{3d}$ & 38.8 GeV \\
TASSO$^{2a}$	& 43.6 GeV 	& $pp$ SMF at CERN$^{3b}$ & 52.6 GeV \\
ALEPH$^{2c}$ 	&   91 GeV 	& $pp$ SMF at CERN$^{3b}$ & 62.2 GeV \\
DELPHI$^{2d}$ 	&   91 GeV 	& $\pbar p$ UA5$^{3e}$ & 200 GeV \\
L3$^{2e}$      	&   91 GeV 	& $\pbar p$ UA5$^{3f}$ & 546 GeV \\
OPAL$^{2f}$    	&   91 GeV 	& $\pbar p$ UA5$^{3e}$ & 900 GeV \\
% & & & \\
\hline
\end{tabular}
\end{center}
\end{table}
distribution
in the Leading Logarithmic Approximation (LLA)$^{1c}$;
some additional
oscillations are predicted by the Leading Double Logarithmic
Approximation (DLA), the corresponding ratios
still being, however, positive
and globally decreasing$^{1c}$.
Predictions of a minimum of $H_{q}$ at
$q_{min_{1}}\simeq 4$ and indications of a further minimum
at $q_{min_{2}} \simeq 2 q_{min_{1}}$ are instead given
by next-to-next to leading order gluodynamics
when the non-linear terms in the equations
for the generating functions of multiplicity
distributions are properly treated$^{1d}$.
In this paper some results$^{1e,1f}$ on the behaviour of the ratio $H_{q}$ for
charged multiplicity from inelastic collision data  are reported.

\vspace{1pt}
\begin{flushleft}
{\bf \large 2. Experimental Results}
\end{flushleft}
\vspace{1pt}

Here we consider charged multiplicity data in full phase-space
from $e^{+}e^{-}$ annihila- tions$^{2}$
in the energy range $\sqrt{s}=$ 22 to 91 GeV and from $hh$ collisions$^{3}$
in the energy range $\sqrt{s}=$ 23.8 to 900 GeV.
The experiments are listed in Table 1.
The results, up to the 16-th order, are given
in fig.s 1 for $e^{+}e^{-}$ experiments and in fig.s 2 for $hh$
experiments; in each figure the experiments are sorted
from top to bottom by increasing energy.
The solid line interpolated to the points is meant only to guide the eye,
for a better reading of the figures.

Most $e^{+}e^{-}$ data display, as a function of the rank, an $H_{q}$
behaviour characterized by a steep descent, taking place at the lower moment
ranks, followed by a negative minimum, of order $10^{-4}$ and
located between $q=4$ and $q=6$. Then  $H_{q}$ becomes positive,
reaches a maximum and gives a second negative minimum between
$q=9$ and $q=13$. In some of the figures also a second positive maximum is
seen. This particular oscillatory trend is clear in the four LEP experiments,
corresponding to fig.s 1e-1h, while for the TASSO and HRS data,
fig.s 1a-1d show less pronounced and regular oscillations.

Qualitatively similar regularities are observed
in the $hh$ outcomes of fig.s 2, where an exponential descent,
showing positive values, is followed by at least two negative minima.
Here, however, the initial descent is less steep and the
order of magnitude of the minima (between
$10^{-3}$ and $10^{-2}$) is larger than before.

\newpage
\vspace{1pt}
\begin{flushleft}
{\bf \large 3. Conclusions}
\end{flushleft}
\vspace{1pt}

Though we are at a preliminary approach to the whole problem,
it is already clear that the
proposed ratio $H_{q}$ of cumulant and
factorial moments is a sensitive measure of multiplicity
distributions, since it helps in distinguishing various distributions
which are hardly separated on the base of factorial moments only.
The data presented here show that
both in $e^{+}e^{-}$ and $hh$
full phase-space charged multiplicity distributions
and in a wide energy range $H_{q}$
displays, as a function of the rank,
a rather regular  behaviour with at least two negative minima.
If a rough comparison between theory and data is performed one can observe
that, experimental outcomes cannot be accounted
for by NBD, LLA or DLA predictions;
only next-to-next to leading gluodynamics predictions of ref. [1d]
are in qualitative agreement with data.
One could expect such qualitative features to be appropriate for
$e^{+}e^{-}$ annihilation since perturbative
QCD is supposed to hold for
hard processes and high energies, so it is a surprise to observe similar
features also in soft $hh$ collisions which are out of the scope of those
theoretical approximations.
Of course a comparison of the experimental
behaviour of $H_{q}$ with the results of ref [1d] is not proper
on the quantitative level, because quarks, higher-order terms and
confinement have not been considered there. Nonetheless
the occurrence of the same qualitative features in so different interactions
and energies deserves some theoretical attention and further
experimental investigations.

%%%%%%%%%%%%%%%%%%%%%%%%%%%%%%%%%%%%%%%%%%%%%%%%%%%%%%%%%%%%%%%%%%%%%%%%%%

\newpage
\begin{figure}
\label{ALL3}
\vspace{450pt}
%\special{psfile=f3.ps hoffset=-33 voffset=-55 hscale=80 vscale=70}
\caption{$H_{q}$ vs. $q$ for various
$e^{+}e^{-}$ experiments. On the left the first $H_{q}$
orders in logarithmic scale, on the right the orders from
$q$=3,4 to 16.  The experimental outcomes are ordered
according to their energy, that increases from top to bottom:
 (a)  e+e- 22 GeV TASSO Coll.;
 (b)  e+e- 29 GeV, HRS Coll.;
 (c)  e+e- 34.8 GeV TASSO Coll.;
 (d)  e+e- 43.6 GeV TASSO Coll.;
 (e)  e+e- 91 GeV, ALEPH Coll.;
 (f)  e+e- 91 GeV DELPHI Coll.;
 (g)  e+e- 91 GeV OPAL Coll.;
 (h)  e+e- 91 GeV, L3 Coll. (here the scale has been changed
by a factor 2., with respect to the other experiments).
    }
\end{figure}

\begin{figure}
\label{ALL4}
\vspace{450pt}
%\special{psfile=f4.ps hoffset=-30 voffset=-55 hscale=80 vscale=70}
\caption{$H_{q}$ vs. $q$ for various
$hh$ experiments. On the left the first $H_{q}$
orders in logarithmic scale, on the right the orders from
$q$=3/4 to 16 (notice that the scale on the right is different
from the one used in Fig.s 3); the points associated to
exceedingly large error bars have not been plotted for clearness,
only the corresponding portion of spline has been left.
The experimental outcomes are ordered
according to their energy, that increases from top to bottom:
 (a)  pp 300 GeV/c (C.M.S. energy 23.8 GeV) FNAL 30 in. bubble chamber;
 (b)  pp 30.4 GeV SMF det. at the CERN ISR;
 (c)  pp 800 GeV/c (C.M.S. energy 38.8 GeV) E743 FNAL exp.;
 (d)  pp 52.6 GeV SMF det. at the CERN ISR;
 (e)  pp 62.2 GeV SMF det. at the CERN ISR;
 (f)  $\overline{p}p$ 200 GeV UA5 Coll.;
 (g)  $\overline{p}p$ 546 GeV UA5 Coll.;
 (h)  $\overline{p}p$ 900 GeV UA5 Coll. .
  }
\end{figure}
\end{document}